\documentstyle[pre,epsf,aps]{revtex}
\newcommand{\tds}{topological defects}
\newcommand{\gb}{Gay-Berne}
\newcommand{\tone}{type-1}
\newcommand{\tones}{type-1 lines}
\newcommand{\mons}{monopoles}
\newcommand{\disc}{disclination}
\begin{document}
\draft
\title{Phase--ordering dynamics of the \gb\ nematic liquid crystal}
\author{Jeffrey L. Billeter}
\address{Department of Physics, Brown University, Providence, RI 02912}
\author{Alexander M. Smondyrev}
\address{Department of Chemistry, University of North Carolina at Chapel Hill,
Chapel Hill, NC 27599}
\author{George B. Loriot}
\address{Computing and Information Services, Brown University,
Providence, RI 02912}
\author{Robert A. Pelcovits}
\address{Department of Physics, Brown University, Providence, RI 02912}
\date{\today}
\maketitle
\begin{abstract}
Phase--ordering dynamics in nematic liquid crystals has been the subject of
much active
investigation
in recent years in theory, experiments and simulations. With a rapid quench
from the isotropic
 to nematic phase a large number of  topological defects are formed and
dominate the subsequent
 equilibration process. We present here the results of
a molecular dynamics simulation of the \gb\ model of liquid crystals after such
a quench in a
 system with 65536 molecules. Twist disclination lines as well as \tones\ and
\mons\ were
observed.  Evidence of dynamical scaling was found in the behavior of the
spatial correlation
 function and the density of disclination lines. However, the behavior of the
structure factor
 provides a more sensitive measure of scaling, and we observed a crossover from
a defect
dominated regime at small values of the wavevector to a thermal fluctuation
dominated regime at
large wavevector.
\end{abstract}
\pagestyle{headings}
\section{Introduction}
\par
Topological defects formed during quenches from high--temperature equilibrium
phases are of
 interest in a wide variety
of fields from condensed matter physics to cosmology
\cite{bow-chand,chuang-durr,sal-vol,vil-shell}.
Uniaxial nematic liquid crystals are
excellent materials for studying \tds\ because of the variety of defects they
possess
and because of the ease with which they can be studied experimentally. Tables
of
processes involving defects, such as found in \cite{chuang-yurk} are
interesting both from a theoretical and from an experimental point of view.
Simulations
in which actual molecular configurations can be viewed and tracked could
greatly
elucidate these processes and aid our general understanding of defect dynamics
and phase ordering.
This paper represents a step towards these goals.
\par
Simulations of defects in nematics have often used, by analogy with $O(n)$ and
other model
simulations \cite{mond-gold,blund-bray1,nishi-nuki}, a
cell-dynamical scheme \cite{toyoki,zap-gold2,gold1} in which the order
parameter $\psi$
at each site is advanced
in time according to a time-dependent Ginsburg-Landau equation, within the one
elastic constant approximation. Others \cite{bed-wind,liu-muth} have performed
Monte Carlo simulations of a discretized  Frank free
energy, including allowance for
elastic anisotropy and surface anchoring.
Still others \cite{kilian1,pis-rubin,hud-lars} have investigated
specific types of defects or processes by directly creating the appropriate
configurations as initial conditions and then evolving the system. While all of
these
approaches have yielded fruitful results, it would certainly be advantageous
to study defects using more realistic off--lattice models with no prior bias
towards forming any particular defect configurations. In this paper we present
results of a simulation of a quench of the Gay--Berne nematic liquid crystal
\cite{gayberne}, a phenomenological fluid model which mimics the behavior of
ellipsoidal molecules interacting through a combination of attractive and
repulsive forces. This model has proven over the past decade to capture the
essential physical features of real liquid crystals \cite{gb-review}, and it is
an appropriate model for studying the formation of topological defects with an
off--lattice model.
\par
This paper is organized as follows. In the next section we review the
classification
of nematic defects, the dynamical scaling hypothesis and the scaling forms of
the real--space correlation function and structure factor. In Sec. III we
present the computational details of our simulation, followed in Sec. IV by a
description of our defect-finding algorithms. Our results and a comparison with
theoretical predictions is presented in Sec. V, which is followed in the last
section by some concluding remarks.
\section{Theoretical Background}
\label{sec:theory}
\par
\begin{figure}
\center
\epsfxsize=12cm
\epsffile{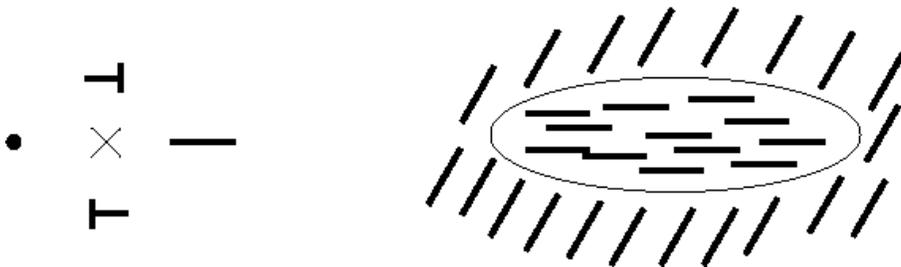}
\caption{(a) Director configuration around a twist disclination line (pointing
out
of the page). On the left of the defect, the director is pointing out of the
page, parallel
to the \disc\ line, while on the right, the director lies in the plane of the
page. Above
and below the defect, the directors are depicted in intermediate orientations.
(b) Illustration of the difference in director orientation between
regions interior and exterior to the twist disclination line. The two regions
have uniform orientations but are rotated with respect to each other by
$90^{\circ}$
along an axis perpendicular to the loop. }
\label{twistdiscfig}
\end{figure}

A basic understanding of the defects in nematics goes back as far as the early
work of
Lehmann \cite{lehm}, but the first quantitative classification was given by
Oseen\cite{oseen}. Topological defect solutions are local minima of  the Frank
free energy \begin{equation} \label{frank}
F = \frac{1}{2}
\int d^3x \left[K_{11}\left({\bf \nabla \cdot \hat{n}}\right)^2 + K_{22}
\left({\bf \hat{n} \cdot \nabla \times \hat{n}}\right)^2 +
K_{33}\left|{\bf \hat{n} \times
\left(\nabla \times \hat{n}\right)}\right|^2\right], \end{equation}
where $\bf \hat {n}$ is the nematic director. A three--dimensional  uniaxial
nematic has stable point (monopole) and line (disclination) defects . The
former include both radial (charge +1) and hyperbolic (charge -1) geometries
\cite{lubensky-pettey} which are topologically equivalent. Disclination line
defects are either of the wedge or twist variety. Either variety is
characterized by a $\pm180^\circ$ director rotation about the line (i.e. the
defects have charge $\pm1/2$). A twist disclination loop is shown in Fig.
\ref{twistdiscfig}. Note that the loop carries zero monopole charge and the
director configuration is uniform at large distances from the loop.
Wedge disclination loops on the other hand carry a net charge, and at large
distances from them the director configuration is equivalent to that of a
monopole of charge 1.
\par
\begin{figure}\center\epsfxsize=12cm \epsffile{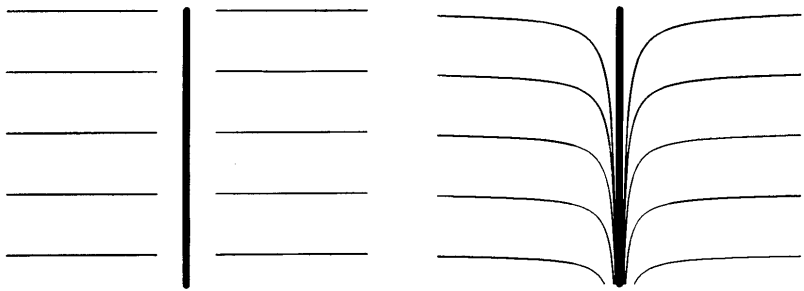}
\caption{(a) Side view of a singular but unstable line with topological charge
+1 (\tone line). (b) Escaped \tone line with
directors tilted into a non-singular configuration.}
\label{escapefig}
\end{figure}

 Another type of line defect
is characterized by director rotations of $\pm360^\circ$ (\tone\ line)and is
unstable to  ``escaping in the third dimension,'' into a
non-singular configuration \cite{klembk} (see Fig. \ref{escapefig}). These
escaped structures
can still
be observed experimentally \cite{meyer1,parg-men,will-clad}, however, and we
can also
visualize them in our simulation.

\par
The dynamical scaling hypothesis \cite{bray1} for phase ordering processes
asserts that there is a characteristic length $L(t)$ (\textit{e.g.,} the domain
size
or defect separation) such that the system appears to be time-independent (in a
statistical sense) when all lengths are rescaled by $L(t)$. For nematics,
 theory \cite{bray1} predicts
that $L(t) \sim t^{1/2}$, where $t$ is the time since the ordering process
began (\textit{e.g.,} the time since a temperature quench which leads to an
isotropic--nematic transition). The \disc\ line density $\rho_{disc}$ (total
length of \disc\ lines per unit
volume) should then scale as $L(t)/(L(t))^3 \sim (L(t))^{-2} \sim t^{-1}$ and
the monopole density $\rho_{monop}$ (number
of monopoles per unit volume) as $(L(t))^{-3} \sim t^{-3/2}$. Note that defects
occur at the intersections of domains growing with differing director
orientations (the
Kibble mechanism \cite{kibble}). Until the domains are large enough,
defects are neither well-defined nor well-separated (one needs a defect
separation
larger than the defect core size \cite{gold1}) so
scaling is only assumed to hold at ``late'' times.
Experiments are generally consistent with scaling predictions except perhaps in
the
behavior of monopoles \cite{chuang-yurk,hindmarsh,nag-hot}. Simulations also
demonstrate scaling \cite{toyoki,zap-gold2,liu-muth} but with calculated
exponents somewhat different from theory and experiment. There have also been
indications
in simulations that more than one characteristic length may be present
\cite{zap-gold2},
but this is possibly just a finite-size effect.
\par
The real-space order parameter correlation function $C({\bf r},t)$ and its
Fourier transform
$S({\bf k}, t)$ are widely used probes \cite{bray1} of domain structure and
dynamical scaling. For the nematic order
parameter
\begin{equation} \label{ordparamtens}
Q_{\alpha\beta}({\bf x}) = \frac{3}{2}\left[
{\bf \hat{u}}_\alpha({\bf x}){\bf \hat{u}}_\beta({\bf x}) -
\frac{1}{3}\delta_{\alpha\beta}\right],
\end{equation}
one has the definitions \begin{mathletters} \begin{equation} \label{corrfcn}
C({\bf r},t) = \frac{\int d^dx\ Q_{\alpha\beta}({\bf x},t)Q_{\beta\alpha}({\bf
x}+{\bf r},t)
}{\int d^dx\ Q_{\alpha\beta}({\bf x},t)Q_{\beta\alpha}({\bf
x},t)},\end{equation}
\begin{equation} \label{strucfac}
S({\bf k},t) =
\int d^dr\ e^{i{\bf k \cdot r}} C({\bf r},t) = \frac{Q_{\alpha\beta}({\bf
k},t)Q_{\beta\alpha}(
{\bf -k},t)}{\int d^dx\ Q_{\alpha\beta}({\bf x},t)Q_{\beta\alpha}({\bf
x},t)}.\end{equation} \end{mathletters}
According to the scaling hypothesis \cite{bray1},
the data for the orientationally averaged $C(r,t)$ at different times should
collapse
to a single curve
when distances at time $t$ are rescaled by $L(t)$. Similarly, $S(k,t)$ should
have a
single, underlying scaling form: that is, \begin{mathletters}
\begin{equation} \label{corrscaling} C(r,t) = f\left[r/L(t)\right],
\end{equation}
\begin{equation}\label{strucscaling} S(k,t) =
L^dg\left[kL(t)\right].\end{equation} \end{mathletters}
\par
The late-time behavior of $S(k)$ is determined by the
type and number of defects present in the system. For nematics, $S(k)$ can be
written in the
form \cite{zap-gold1} \begin{equation}\label{strucdefect} S(k,t) =
\rho_{monop}\frac{36\pi^4}{k^6} +
\rho_{disc}\frac{
3\pi^3}{k^5} + \frac{3}{2}\frac{k_BT}{Kk^2},\end{equation} where the
right--hand side includes contributions from the monopoles, disclinations and
thermal fluctuations respectively (the nonsingular \tones\ do not make a
power--law contribution to the structure factor). Here $K$ is the elastic
constant in
the one-constant approximation. For thin nematic films (i.e. two spatial
dimensions, but with a three--dimensional director $\bf \hat {n}$) the monopole
and disclination contributions to eqn. (\ref{strucdefect}) are replaced by a
single contribution proportional to $k^{-4}$ arising from disclination points
characterized by $\pm 180^\circ$ director rotations.  In the three-dimensional
case the disclination contribution proportional to $k^{-5}$ appears for twist
and wedge disclination loops (as well as any curved disclination loop segment)
at wavevectors $k \gg R^{-1}$, where $R$ is the radius of the loop
\cite{zap-gold1}. For smaller wavevectors, there is no power--law contribution
to the structure factor from the twist loops (recall that the director
configuration is homogeneous at large distances from the loop), and wedge loops
contribute a term of the same form as the monopoles.
The defect contributions to the structure factor above are specific examples of
Porod's Law \cite{bray1} which states that
\begin{equation} S(k) \sim \rho\frac{1}{
k^{2d-D}},\end{equation} where $\rho$ is the
defect density, $d$ is the number of spatial dimensions and $D$ is the defect
dimensionality (\textit{e.g.,} points have $D=0$ and lines have $D=1$).
Experiments \cite{yurk-parg} show good
scaling of $S(k)$ with an asymptotic exponent approximately equal to 5, with
the approach through effective exponents lying between 5 and 6
\cite{bray-puri}. Zapotocky and Goldbart \cite{zap-gold1} have shown that such
behavior would be consistent with the presence of sufficient numbers of
monopoles or wedge disclination loops. However, experimentally the population
of monopoles seems too low, and wedge disclinations are energetically less
preferable \cite{anisimov,chandra-rang} than twist disclinations for typical
values of the nematic elastic constants  (though the former defects might be
generated dynamically). From Eq. (\ref{strucdefect}) we see that thermal
fluctuations will dominate the structure factor for sufficiently large
wavevectors satisfying $(\frac{1}{2}k_BT/K \rho_{disc})k^2 > 1$ (assuming that
monopoles are not present in large numbers). This behavior has not been seen in
the experimental studies \cite{wong1,wong2} carried out thus far. As discussed
in \cite{zap-gold1}, the scattering experiments were performed over a time
range where the defect density is sufficiently large that the crossover to
thermal regime is not evident for wavevectors in the visible range.
\section{Simulation Details}
\begin{figure}\center
\epsfxsize=12cm \epsffile{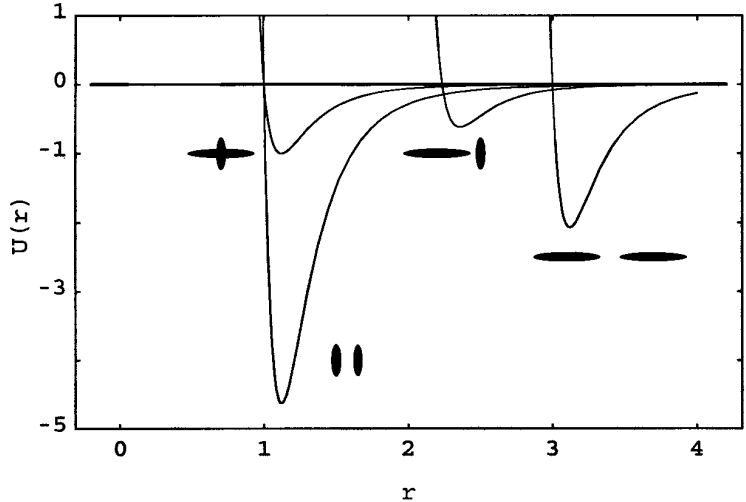}
\caption{Potential energy curves for the Gay-Berne model using the parameters
cited in the text. Curves are shown for four sample molecular pair
configurations as
indicated.}
\label{gbfig}
\end{figure}

We performed a molecular dynamics (MD) simulation using the
Gay-Berne model\cite{gayberne}, an
intermolecular potential similar to the simple Lennard-Jones potential but
extended
to model the anisotropic mesogen shape.
The complete Gay-Berne potential is as follows \cite{luck1}:
\begin{displaymath} U\left({\bf
\hat{u}_i,\hat{u}_j,\hat{r}}\right)=4\varepsilon
\left({\bf \hat{u}_i,\hat{u}_j,\hat{r}}\right) \end{displaymath}
\begin{equation}
\times\left[\left\{\frac{\sigma_0}{r-\sigma\left({\bf
\hat{u}_i,\hat{u}_j,\hat{r}}
\right)+\sigma_0}\right\}^{12}\!-\left\{\frac{\sigma_0}
{r-\sigma\left({\bf
\hat{u}_i,\hat{u}_j,\hat{r}}\right)+\sigma_0}\right\}^6\right],
\label{GBpot}\end{equation} where ${\bf \hat{u}_i,\hat{u}_j}$
give the orientations of the long axes of molecules $i$ and $j$, respectively,
and ${\bf r}$ is the intermolecular
vector (${\bf r}={\bf r_i}-{\bf r_j}$).
The parameter $\sigma\left({\bf \hat{u}_i,\hat{u}_j,\hat{r}}\right)$ is the
intermolecular separation at which the potential vanishes, and thus represents
the
shape of the molecules. Its explicit form is
\begin{eqnarray} \sigma\left({\bf \hat{u}_i,\hat{u}_j,\hat{r}}\right)&=&
\sigma_0\left[1-\frac{1}{2}\chi\left\{
\frac{\left({\bf \hat{r}\cdot\hat{u}_i}+{\bf \hat{r}\cdot\hat{u}_j}\right)^2}{
1+\chi\left({\bf \hat{u}_i\cdot\hat{u}_j}\right)}\right.\right. \nonumber\\
&&\left.\left.{}+
\frac{\left({\bf \hat{r}\cdot\hat{u}_i}-{\bf \hat{r}\cdot\hat{u}_j}\right)^2}{
1-\chi\left({\bf \hat{u}_i\cdot\hat{u}_j}\right)}\right\}\right]^{-1/2},
\end{eqnarray}
where $\sigma_0=\sigma_s$ (defined below) and $\chi$ is
\begin{equation} \chi=\left\{\left(\sigma_e/\sigma_s\right)^2-1\right\}/
\left\{\left(\sigma_e/\sigma_s\right)^2+1\right\}. \end{equation} Here
$\sigma_e$
is the separation between two molecules when they are oriented end-to-end, and
$\sigma_s$ the separation when side-by-side.
The well depth $\varepsilon\left({\bf \hat{u}_i,\hat{u}_j,\hat{r}}\right)$,
representing the anisotropy of the attractive interactions, is
written as \begin{equation}\varepsilon\left({\bf
\hat{u}_i,\hat{u}_j,\hat{r}}\right)
=\varepsilon_0\varepsilon^\nu\left({\bf
\hat{u}_i,\hat{u}_j}\right)\varepsilon'^{\mu}
\left({\bf \hat{u}_i,\hat{u}_j,\hat{r}}\right),\end{equation} where
\begin{equation}\varepsilon\left({\bf \hat{u}_i,\hat{u}_j}\right)=\left\{1-
\chi^2\left({\bf
\hat{u}_i\cdot\hat{u}_j}\right)^2\right\}^{-1/2},\end{equation} and
\begin{eqnarray} \varepsilon'\left({\bf \hat{u}_i,\hat{u}_j,\hat{r}}\right)&=&
1-\frac{1}{2}\chi'\left\{
\frac{\left({\bf \hat{r}\cdot\hat{u}_i}+{\bf \hat{r}\cdot\hat{u}_j}\right)^2}{
1+\chi'\left({\bf \hat{u}_i\cdot\hat{u}_j}\right)}\right. \nonumber\\
&&\left.{}+
\frac{\left({\bf \hat{r}\cdot\hat{u}_i}-{\bf \hat{r}\cdot\hat{u}_j}\right)^2}{
1-\chi'\left({\bf \hat{u}_i\cdot\hat{u}_j}\right)}\right\}, \end{eqnarray}
with $\chi'$ defined in terms of $\varepsilon_e$ and $\varepsilon_s$, the
end-to-end
and side-by-side well depths, respectively, as
\begin{equation}
\chi'=\left\{1-\left(\varepsilon_e/\varepsilon_s\right)^{1/\mu}\right\}/
\left\{1+\left(\varepsilon_e/\varepsilon_s\right)^{1/\mu}\right\}.
\end{equation}
The overall energy scale is set by the value of $\varepsilon_0$. Some
representative
plots of the \gb\ potential energy curves are shown in Fig. \ref{gbfig}.
\par
For the adjustable parameters, we used the values suggested by Berardi et al.
\cite{ber-emer}: $\mu=1,\nu=3,\sigma_e/\sigma_s=3$, and
$\varepsilon_s/\varepsilon_e=5$. These values yield a nematic phase over a
wider range of temperatures, compared to the original parameterization chosen
by Gay and Berne
\cite{gayberne}.
The temperature was controlled by velocity rescaling \cite{allen-tild}, and the
density was
fixed at $\rho^*=0.3$ with the dimensions of the simulation box in the ratio
2:2:1.
Periodic boundary conditions were applied. The system was equilibrated
at $T^*=3.6$ in the isotropic phase for 130000 MD time steps (with a
dimensionless time step of 0.004) and then a quench to
$T^*=3.2$ was implemented (the nematic-isotropic transition temperature is
approximately 3.5). The system gradually came to equilibrium in the nematic
phase with the order parameter $S$ (the largest eigenvalue of
$Q_{\alpha\beta}$) saturating at a value of $0.69$ over the next 100000 steps.
We used a
domain decomposition  approach on the Cray T3E at the San Diego Supercomputing
Center.
Briefly, the domain decomposition approach involves dividing up the simulation
volume into a number of
cells, each controlled by a different processor (we used 64 cells). Because the
\gb\
potential is short-ranged, most of the intermolecular interactions involve
same-cell
molecules; thus, only the relatively small number of molecules near cell
boundaries will
require interprocessor communications. The cell scheme and the required
communications
are somewhat difficult to implement, but provide very significant computational
speedups.
A computational scheme similar to ours
is described in more detail in \cite{wils-allen1} (although we used a slightly
different
communication scheme in which an additional map tracking the specific subcells
to be
transferred between specific neighbors was implemented). We obtained timings
virtually
identical to those reported in the latter reference (on the order of 1 second
per timestep with a 64 node partition of the Cray). Our computation time
increases linearly with the number of particles and with the number of
processing elements, indicating good scalability of our code.
\section{Defect-Finding Methods}
\label{sec:algorithm}
A preliminary step for locating defects is to break the system into a lattice
of cubic bins.
Note that the creation of a lattice is strictly for convenience in defect
finding; the time
evolution of the system allows for complete translational freedom. Even
experiments,
of course, have ``binning'' inherent in the resolution of the optical
microscopy. Within each bin, the order parameter
tensor $Q_{\alpha\beta}$, Eq. (\ref{ordparamtens}), was calculated, its largest
eigenvalue taken as the local
order parameter $S$ and the corresponding eigenvalue as the local director
${\bf \hat{n}}$.
The bin size was chosen so that the core size of the \disc s, determined from
the distance
over which $S$ dropped significantly below the background value, was of the
order of one lattice
spacing. In our case, this resulted in a $16 \times 16 \times 8$ lattice, each
bin
holding roughly 30 molecules. It is with this lattice of orientation vectors
that we began
analyzing planes parallel to the $x$, $y$ and $z$ axes for the presence of
defects.
Note that while it is convenient to work with the
orientation \textit{vectors} on the lattice, we must remember that the actual
directors
are \textit{headless} (expressing the symmetry upon rotation by 180$^{\circ}$
about an axis
perpendicular to the director) and so some care must be taken to account for
this.
\par
\begin{figure}\center
\epsfxsize=12cm \epsffile{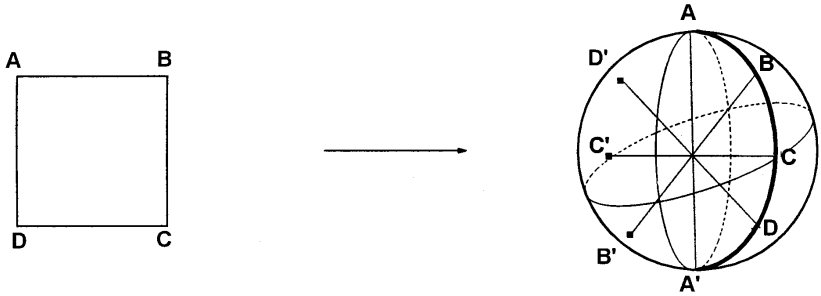}
\caption{The disclination-finding algorithm. The directors at the corners of
the
lattice cube face shown on the left are tracked on the order parameter space
sphere shown on the right. The diameters
$AA^\prime,BB^\prime,CC^\prime,DD^\prime$ correspond to the axes of the
headless director at the real--space lattice sites $A,B,C,D$ respectively.}
\label{discalgfig}
\end{figure}

A nice method of searching for \disc s was introduced in \cite{zap-gold2} (see
also
\cite{strobl}).
Consider the directors at the corners of a square (one of the faces of a cube)
in
our three--dimensional lattice. The idea is to track the course of the
intersections of these
vectors with the order parameter sphere (actually the projective plane $RP_2$)
as one
moves around the corners of the real-space square (Fig. \ref{discalgfig}).
Starting with the
intersection of ${\bf \hat{n}}_A$ with the sphere, one then takes as the next
point either
the intersection of ${\bf \hat{n}}_B$ or $-{\bf \hat{n}}_B$, whichever is
closest
to ${\bf \hat{n}}_A$'s
intersection. Once this point is determined, either ${\bf \hat{n}}_C$ or
$-{\bf \hat{n}}_C$ is
used, depending on the proximity to the previously defined point, and so on.
Once the last
point (from corner D) is determined, one looks at whether its intersection is
in the same
hemisphere as the starting point's. If so, no defect is present---the path in
order
parameter space is deformable to a single point, \textit{i.e.,} a uniform
configuration.
If the first and last points are in different hemispheres, however, then a
\disc\ line
is taken to cut through the center of the square and is oriented perpendicular
to the plane of
the square.
\par
To find escaped \tones\, we performed a similar procedure, except that we
measured the
actual arclength swept out as one moves from each intersection to the next.
Arclengths
greater than $\pi$ are counted as \tones\ if the lattice square has not already
been
determined to hold a \disc\ (obviously, there is some overlap in the methods).
The
arclength $\pi$ corresponds to an escaped structure similar to that of Fig.
\ref{escapefig}
with an opening angle of about 30$^{\circ}$. Experimentally, smaller opening
angles
are observable as \tones, but smaller cutoff values of the arclength produced
too many
random \tone\ line segments unconnected with each other or with \disc s, a
situation not in accord with experiments.

\par
Finally, to look for \mons, we used the method from \cite{toyoki}. Each of the
six faces of a lattice cube is divided into two triangles and the corresponding
directors
are used to map out spherical triangles on the order parameter sphere. Total
areas on the
sphere greater than $2\pi$ are considered to be \mons\ and the defect is placed
at
the center of the corresponding lattice cube.
Note that in all these defect-finding procedures,
one must be careful to apply periodic boundary conditions to the edge lattice
sites.
\par
\begin{figure}\center
\epsfxsize=12cm \epsffile{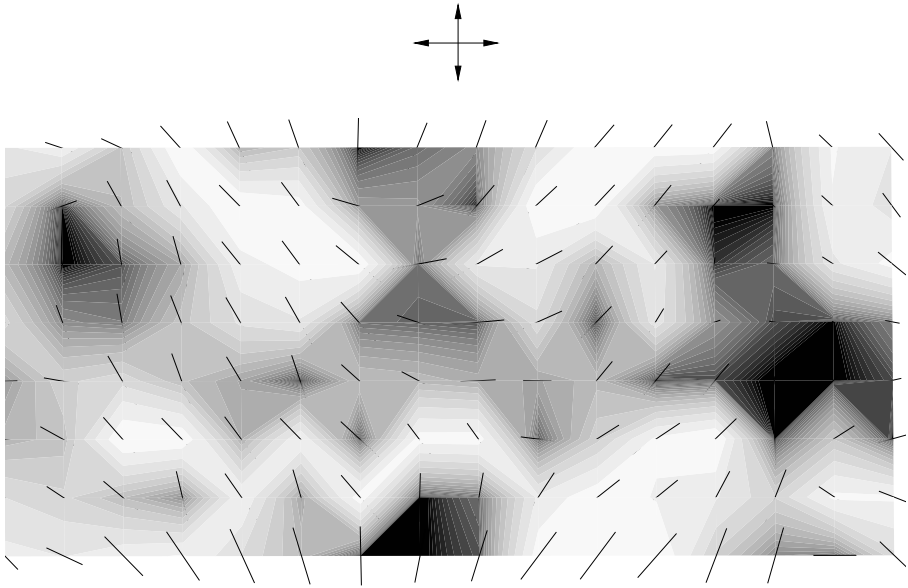}
\caption{Simulated crossed-polarizer image with actual director configuration
superimposed. The crossed-polarizer image is the result of applying the
M\"{u}ller matrix method to the single lattice plane shown. The
disclination line (with topological charge 1/2) at the top-center of the image
is clearly indicated by two brushes.}
\label{xpolarfig}
\end{figure}

One could also consider simulating the effect of crossed polarizers on
individual planes.
We used the method of \cite{ondris-crawford,schell} which, phrased in the
language of the Stokes
parameters, uses  M\"{u}ller matrices to
simulate the effect of a group of molecules on the polarization of incoming
light.
In this method, one must set values for the ordinary and extraordinary
refractive indices;
we used typical experimental values between 1.5 and 2 \cite{degennes}. The
remaining
free parameter, the ratio of the thickness of the cell to the wavelength of
light, was chosen to be the value which makes the calculated outgoing intensity
for
molecules oriented at 45$^{\circ}$ to the crossed polarizer directions equal to
one; we used a
value of 2.5. Visualizing the resulting contour plots is aided by choosing an
exponential
distribution of contour values in order to sharpen the dark areas (the
``brushes''
\cite{degennes}). This
method did yield planes in which two clear brushes met at fairly well-localized
points (Fig. \ref{xpolarfig}),
indicating the presence of a \disc\, but in general, the brushes and
intersections were simply not well-determined enough to be useful.
We estimate that an order of magnitude increase in the number of
bins  (corresponding to several
million \gb\ particles) would be required to use this crossed polarizer
approach quantitatively.
\par
\section{Results}
\subsection{Coarsening sequence}
\label{sec:coarsening}
\begin{figure}\center
\epsfxsize=12cm \epsffile{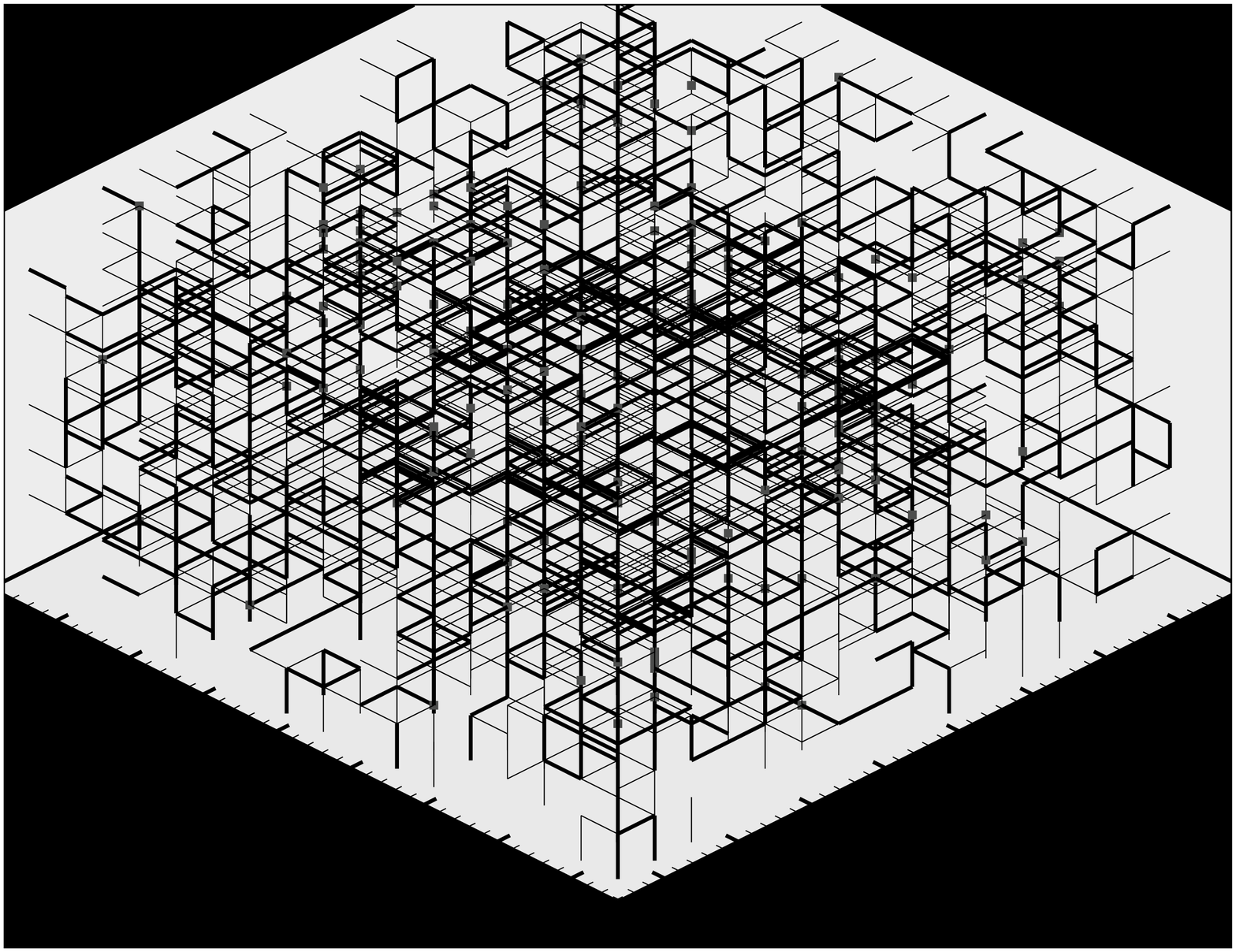}
\epsfxsize=12cm \epsffile{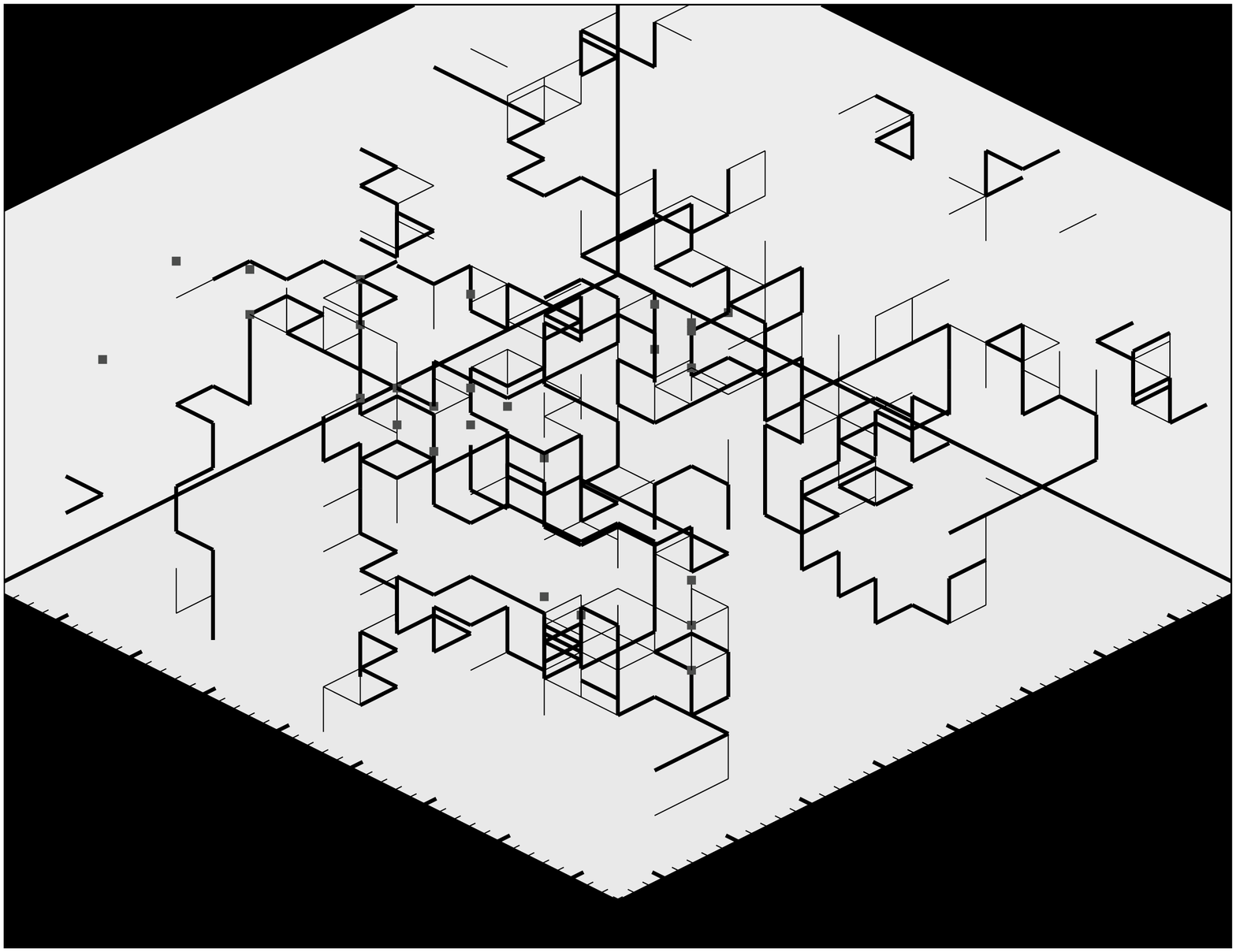}
\epsfxsize=12cm \epsffile{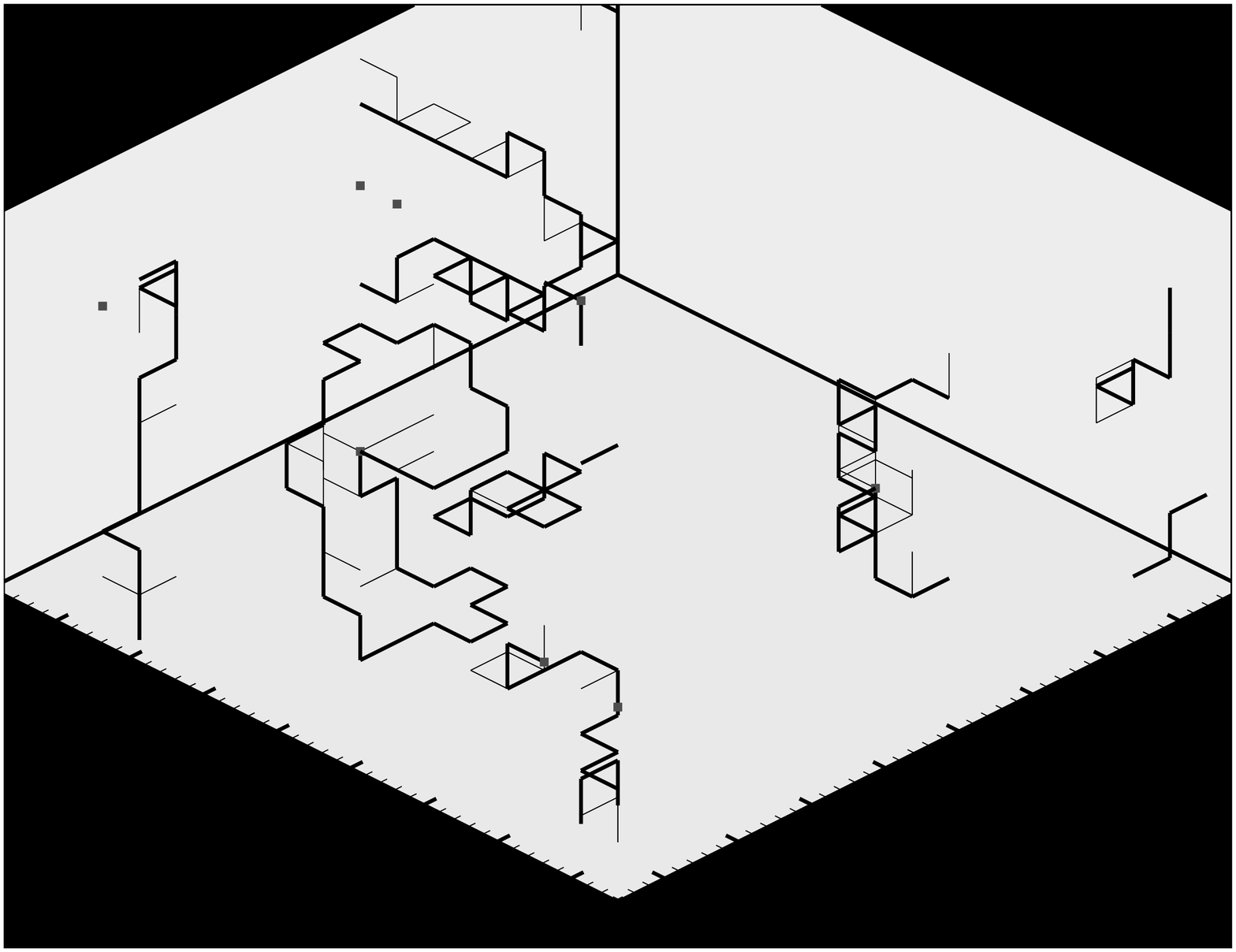}
\epsfxsize=12cm \epsffile{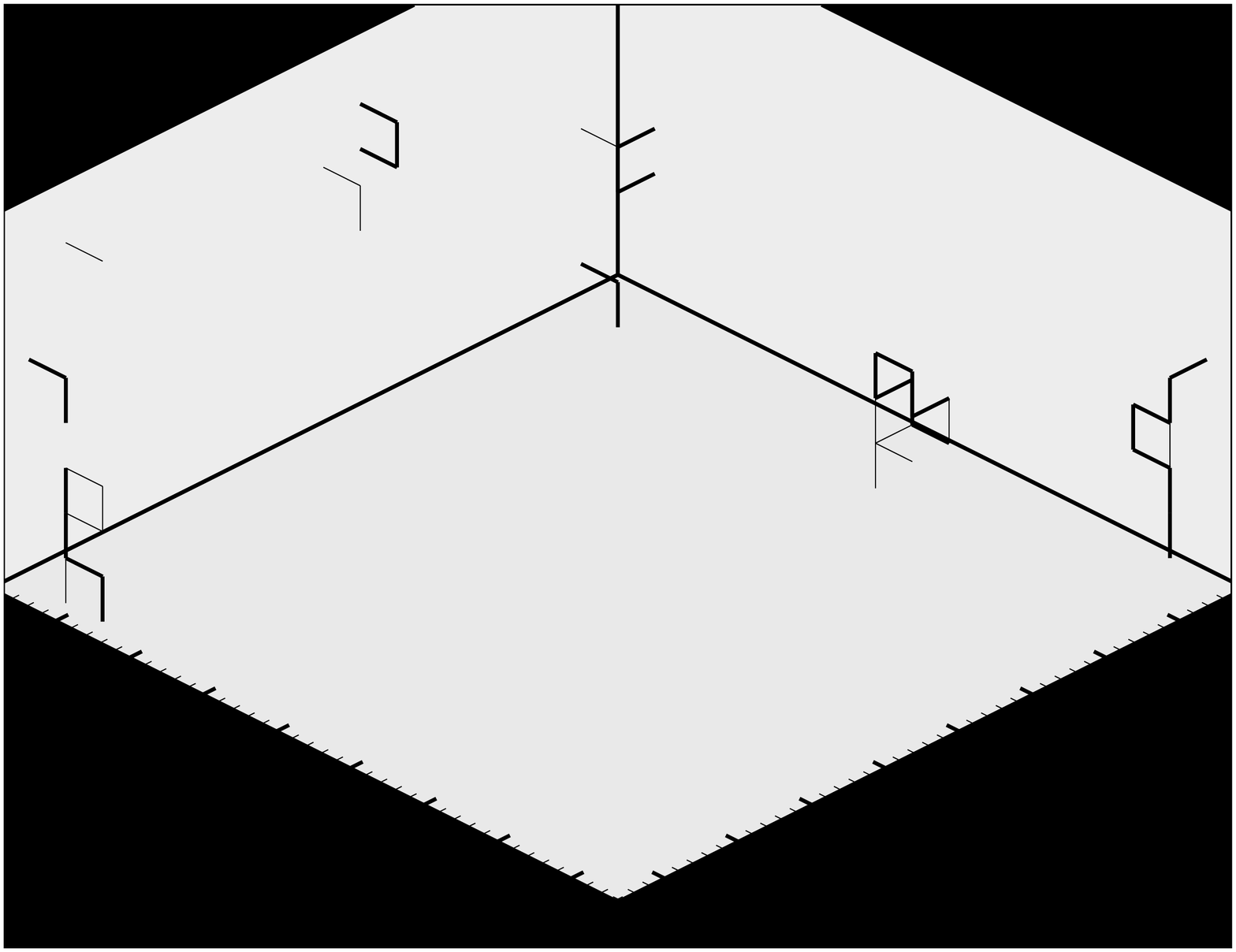}
\caption{Coarsening sequence at times (a) $t=2$, (b) $t= 14$, (c) $t=25$, and
(d) $t=37$, with $t=0$ corresponding to the instantaneous temperature quench
from the isotropic to the nematic phase.  Filled squares represent point
defects,
thin lines represent type-1 lines and thick lines (for emphasis) represent
disclinations. Note that with periodic boundary conditions, all disclination
lines form closed loops. An animation of the coarsening sequence is available
on our web site \protect \cite{web}.}
\label{coarsefig}
\end{figure}

\begin{figure}\center
\epsfxsize=12cm \epsffile{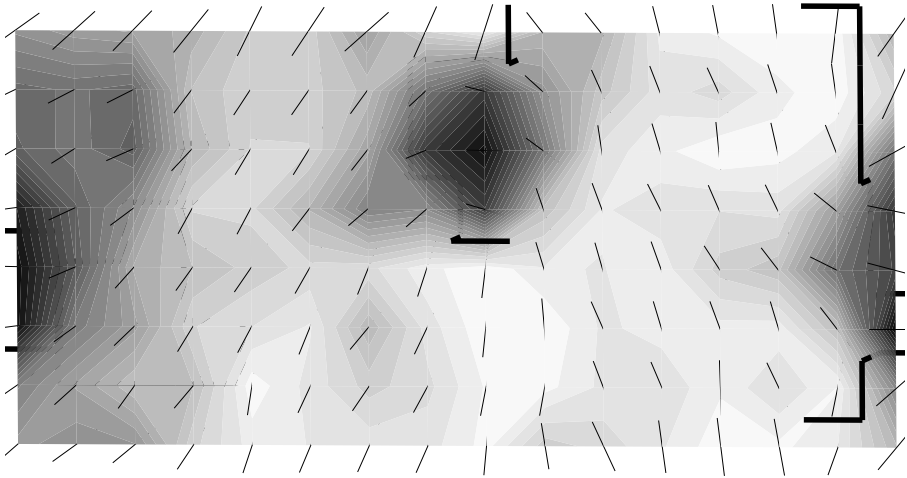}
\caption{Indication of a twist disclination. A single lattice plane of
directors
is shown with disclinations indicated as thick lines. Dark areas indicate local
directors perpendicular to the global director (along the vertical axis) while
light
areas indicate parallel orientations. The dark region in the center of the
figure falls inside a disclination loop, clearly indicating a twist
disclination.
Compare with Fig. \ref{twistdiscfig}(b).}
\label{twistfig}
\end{figure}

With the methods described, we observed a coarsening sequence---compare
Fig. \ref{coarsefig} with similar figures in \cite{chuang-yurk}--which
exhibited
most of the general behaviors observed experimentally \cite{chuang-yurk}. An
animation of our results is available on our web site \cite{web}.
Shortly after the quench, there was a dense tangle of defect lines. This tangle
gradually
thinned out and we could clearly identify and follow individual defect loops.
With the exception of one wedge disclination line \cite{bouligand} running
through the sample, all of the
\disc\ lines were of twist type (see Fig. \ref{twistfig}) and with periodic
boundary conditions formed closed loops.  The presence of twist lines is
consistent \cite{anisimov,chandra-rang} with the relative values of the elastic
constants in the Gay-Berne nematic \cite{Allen-elastic}, namely
$K_2<(K_1+K_3)/2$. Apart from the exception mentioned above we saw no evidence
for dynamically generated wedge disclination lines that might contribute
substantially to the structure factor. Combination, separation and collapse of
the loops were all observed. The \disc\
loops appeared to experience minimal center-of-mass displacement and were
relatively
long-lived structures. Type-1 lines took the form of single line segments or
small partial
loops virtually always connected to \disc\ line segments and often forming
bridges (much
like the $T$ intersections of \cite{chuang-yurk})
between \disc\ segments from the same or distinct loops. Type-1 lines tended to
fluctuate
on much shorter time scales than \disc s, which seems reasonable given that the
former are not topologically stable. One interesting observation is that
\tones\
often appeared as precursors to or remnants of the motion of \disc\ line
segments. For
example, the appearance of a \tone\ line or several connected lines jutting out
from a \disc\
was often followed by a kink or bend developing in the \disc\ line at that
point.
Similarly, the removal of kinks or bends often left behind \tones\ for some
period of time.
The \tones\ seemed to be initially defining and afterwards retaining a memory
of the
\disc\ path. Also, the emergence of distinct \disc\ loops from localized
tangles often
included the breaking of numerous \tone\ ``bonds'' between the loops.
 \par
\begin{figure}[b]\center
\epsfxsize=10cm \epsffile{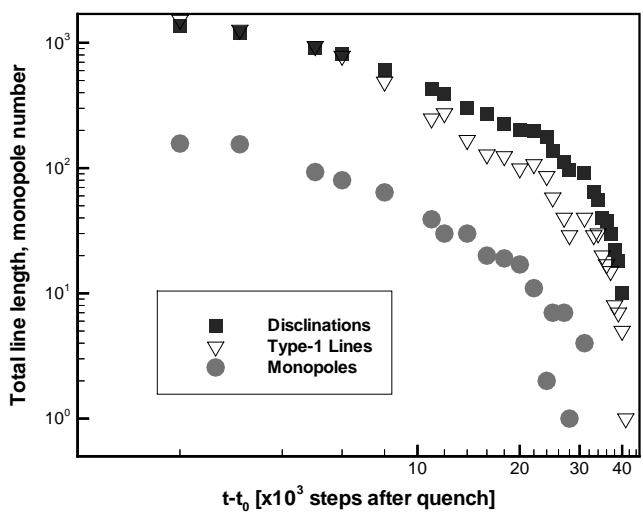}
\caption{Time behavior of the various defect quantities: total length of
disclination lines and type-1 lines and total number of monopoles.}
\label{defects}
\end{figure}
\begin{figure}[t]\center
\epsfxsize=10cm \epsffile{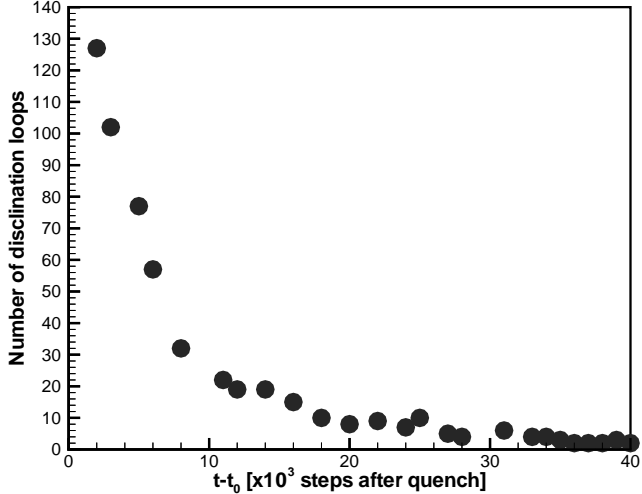}
\caption{Number of disclination loops as a function of time.}
\label{loops}
\end{figure}

The \mons\ we observed fluctuated even more rapidly than the \tones, although
in
many cases, their positions remained constant, on average, over relatively
longer times.
Because of their fluctuations, it is difficult to make any reliable statements
about
specific monopole behaviors such as monopole-antimonopole annihilation, for
example.
We never observed monopole formation upon \disc\ loop collapse, a result
consistent with the presence of only twist \disc\ loops. All of the above
processes are best observed in the animations we provide on our web site
\cite{web}. The total line length of disclination lines and type-1 lines as
well as the number of monopoles is plotted in Fig. \ref{defects} as a function
of time. The total number of disclination loops is plotted as a function of
time in Fig. \ref{loops}.

\subsection{Real--space correlation function $C(r,t)$}
\label{sec:corrfn}
\begin{figure}\center
\epsfxsize=12cm \epsffile{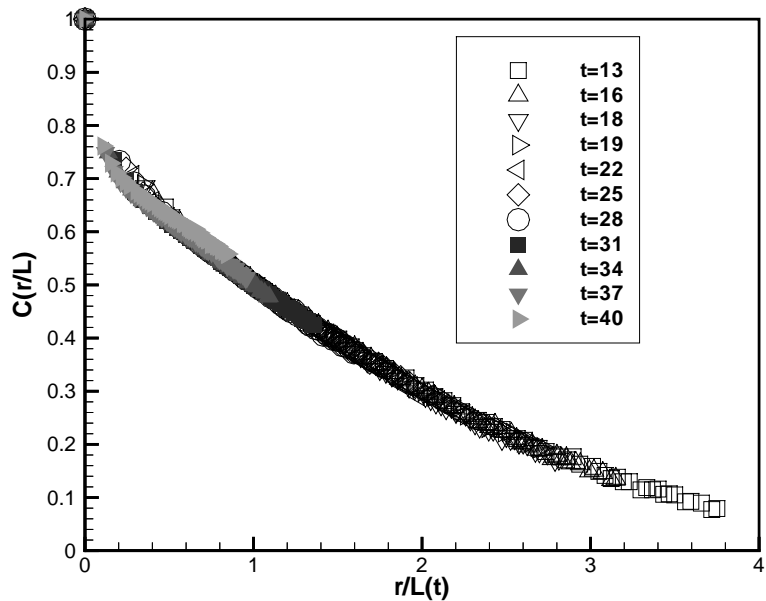}
\epsfxsize=12cm \epsffile{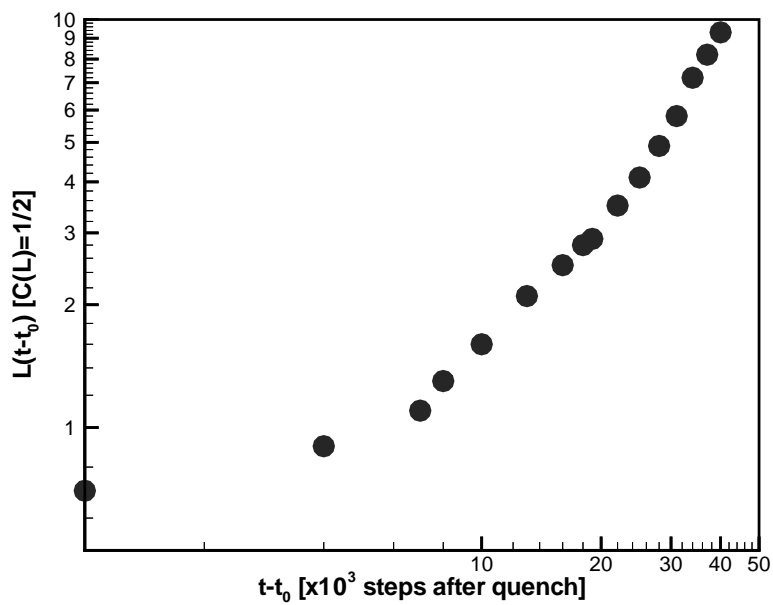}
\caption{Log-log plot of disclination line density (line length per unit
volume) versus the characteristic length $L$ shown in Fig. \ref{corrsfig}(b).
The straight line is a least squares fit with slope $-1.99 \pm 0.23$. The range
of $L$ shown corresponds to the time range $t=13$ to $t=40$ where scaling
behavior of $C(r,t)$ is observed as shown in Fig. \ref{corrsfig}(a).}
\label{scaling}
\end{figure}

\begin{figure}\center
\epsfxsize=10cm \epsffile{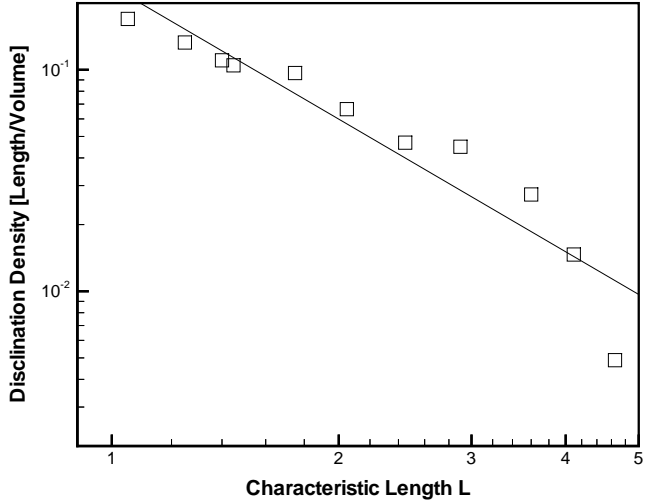}
\caption{a) Correlation function $C(r,t)$  for times ranging from $t=13$ to
$t=40$ with
distances rescaled by the characteristic lengths defined by $C(r=L(t))=1/2$.
 (b) Time behavior of the characteristic length $L(t)$. Note that the data in
this figure was obtained by using a smaller bin size (corresponding to a $32
\times 32 \times 16$ lattice) than the previous figures.}
\label{corrsfig}
\end{figure}

To calculate the correlation function $C(r,t)$, Eq. (\ref{corrfcn}), we reduced
our bin dimensions by a factor of 2 (yielding a $32 \times 32 \times 16$
lattice size) to obtain a larger data
set. We obtained curves for $C(r,t)$ at times spanning
the entire coarsening process. Motivated by the dynamical scaling hypothesis
Eq. (\ref{corrscaling}) we attempted to collapse our data to a single curve
with appropriate rescaling of distances. From $t=13$ (in units of thousands of
steps since
the temperature quench) until $t=40$ when nearly all of the defects
disappeared, the $C(r,t)$ curves for different times collapse to
a single curve (Fig. \ref{corrsfig} (a)) upon rescaling distances by a length
scale $L(t)$ chosen so that $C(r=L(t),t)=1/2$. This particular choice $L(t)$
for the characteristic length scale was first suggested in \cite{zap-gold2},
and is the most accurate to implement numerically. The time dependence of
$L(t)$ is shown in Fig. \ref{corrsfig}(b). Our system is not large enough to
extract a reliable power--law for the growth of $L(t)$. However, according to
the dynamic scaling hypothesis the length $L(t)$ defined by the above criterion
should differ at most by prefactors or subdominant contributions at late times
from other characteristic length scales of the system. For example, as we noted
in Sec. \ref{sec:theory} the disclination line density should scale as
$(L(t))^{-2}$. In Fig. \ref{scaling} we plot the disclination line density as a
function of $L(t)$, over the range of times ($t=13$ to $t=40$) where we found
good scaling of $C(r,t)$. A least squares fit yields an exponent $1.99 \pm
0.23$, {\it consistent} with dynamical scaling. However, the range of times
over which coarsening occurs is too limited (due to the small system size) to
allow us to fully assess the validity of the dynamical scaling hypothesis.
Similarly, while the collapse of the correlation function data to a single
scaling curve is consistent with the predictions of dynamical scaling, the
range of distances and times is too limited to provide more definitive support
for the hypothesis. Furthermore, as we discuss in the next section when we
examine the structure factor dynamical scaling may in fact be breaking down in
the range $(r/L(t))<1$, even though this is not evident from the scale of the
plot of $C(r,t)$.

\subsection{Structure factor $S(k,t)$}
\begin{figure}\center
\epsfxsize=10cm \epsffile{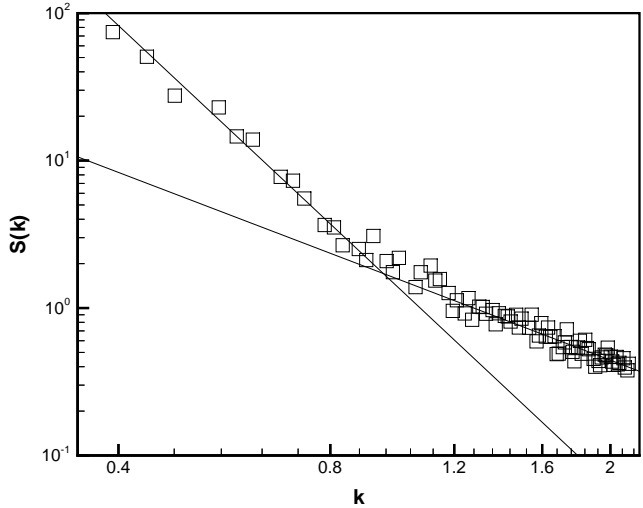}
\caption{Structure factor as a function of $k$ at time $t=27$. The exponents of
the power--law fits at small and large $k$ are 4.3 and 1.8 respectively. At
small $k$ the structure factor is dominated by defects, while at large $k$
thermal fluctuations dominate.}
\label{strucfig27}
\end{figure}

\begin{figure}\center
\epsfxsize=10cm \epsffile{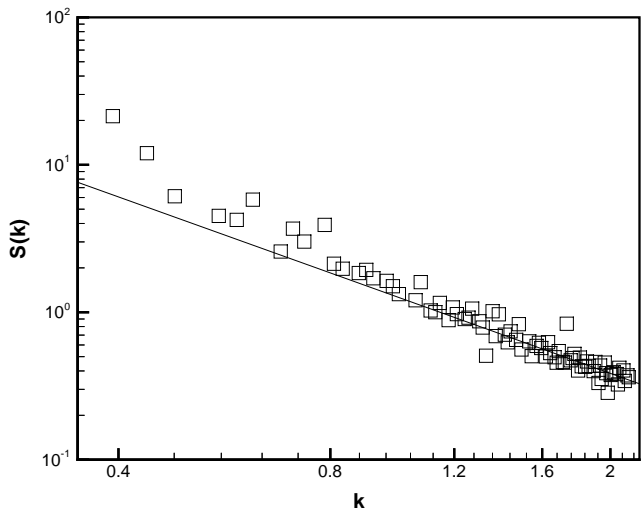}
\caption{Structure factor as a function of $k$ at time $t=40$, near the end of
the coarsening sequence. The crossover between the thermal fluctuation regime
at large $k$ and the defect dominated regime at small $k$ occurs at smaller
values of $k$ than at earlier times in the coarsening sequence (compare with
Fig. \ref{strucfig27}). Due to the relatively small number of data points in
the defect dominated regime, we have not attempted a power--law fit.}
\label{strucfig40}
\end{figure}

\begin{figure}\center

\epsfxsize=10cm \epsffile{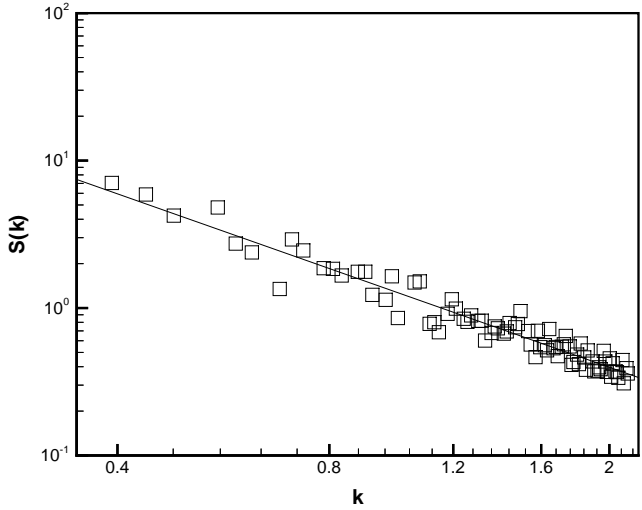}\caption{Structure
factor as a function of $k$ at time $t=70$, after all of the defects have
disappeared. The data is fit with an exponent of 1.7. With the coarsening
process completed, only thermal fluctuations contribute to the structure
factor.}
\label{strucfig70}
\end{figure}

\begin{figure}\center
\epsfxsize=10cm \epsffile{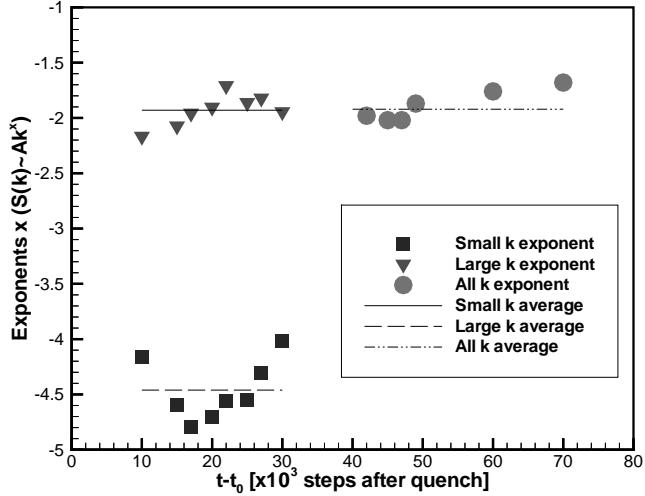}
\caption{Values of the exponents used to fit the structure factor $S(k)$ at
small and large values of $k$ as a function of time. Topological defects
dominate the small $k$ behavior until the defects disappear around $t=40$,
while thermal fluctuations dominate at large $k$. At later times thermal
fluctuations are the only contribution to $S(k)$ for all $k$, and a single
exponent fits the entire range of $k$. During the last stages of the coarsening
sequence (between $t=30$ and $t=40$) we do not have sufficient numbers of data
points to fit the small $k$ behavior because the crossover to the thermal
fluctuation regime occurs at small $k$. However, the large $k$ behavior
continues to be fit well with an exponent of approximately 2 (see Fig. \ref
{strucfig40}). The solid and dashed  lines indicate the average values of the
exponents used to fit the large and small $k$ regimes of $S(k)$ during the
coarsening sequence; the values are 1.9 and 4.5 in these regimes respectively.
The dashed--dotted curve indicates the average exponent, 1.9, that fits all of
the $k$ data after the defects have disappeared.}
\label{exponents}
\end{figure}

We computed the structure factor $S(k,t)$, Eq.(\ref{strucfac}), by first
evaluating the Fourier transform of the nematic order parameter, Eq.
(\ref{ordparamtens}):
\begin{equation} \label{fourierordparam}
Q_{\alpha\beta}({\bf k}) = \frac{V}{N}\sum_i\frac{3}{2}\left[
{\bf \hat{u}}_{i\alpha}{\bf \hat{u}}_{i\beta} - \frac{1}{3}
\delta_{\alpha\beta}\right]\exp(i{\bf k \cdot r}),
\end{equation}
where $V$ and $N$ are the system volume and number of molecules respectively.
As in Sec.  \ref{sec:corrfn} we used a $32 \times 32 \times 16$ lattice. The
wavevectors ${\bf k}$ have components which are multiples of the minimum values
commensurate with the MD cell sizes in each direction. Motivated by the
expression Eq. (\ref{strucscaling}) for the structure factor we plotted our
results in log--log form. Representative results are shown in Figs.
\ref{strucfig27}, \ref{strucfig40} and \ref{strucfig70}, where we have computed
$S(k)$ for values of $k \ge 2 \pi /16$, the minimum commensurate value along
the shortest dimension of the cell, and less than $k \simeq 2$. For values of
$k$ larger than 2 we are unable to fit our data to the long wavelength
expression Eq. (\ref{strucfac}) for $S(k)$. Fig. \ref{strucfig27} corresponds
to time $t=27$ when there are still a sizable number of defects, Fig.
\ref{strucfig40} corresponds to $t=40$ near the end of the coarsening sequence,
while Fig. \ref{strucfig70} corresponds to $t=70$ well beyond the end of the
coarsening sequence. The data in the latter figure can be fit over nearly the
entire range of $k$ by a power law $S(k) \sim k^{-1.9}$, consistent with a
purely thermal fluctuation contribution to the structure factor.  On the other
hand, we note that in the figure corresponding to a time midway through the
coarsening process (Fig. (\ref{strucfig27})), there is an apparent crossover in
the behavior of $S(k)$ as a function of $k$. For small values of $k$, $S(k)$
can be fit to the power law form, $S(k) \sim k^{-4.5}$, while at larger $k$ the
exponent is approximately 1.9. In Fig. \ref{exponents} we show the power--laws
obtained at small and large values of $k$ during most of  the coarsening
sequence and beyond. This crossover behavior is consistent at least in part
with the predictions of \cite{zap-gold1}. At large values of $k$ during the
coarsening process (and at all values of $k$ when the process is complete),
thermal fluctuations dominate and we fit our data with an exponent close to 2.
At small $k$ during the coarsening process our fit yields an exponent of
approximately 4.5, clearly distinct from the thermal fluctuation form. As
indicated in Fig. \ref{defects}, the total disclination line length is in
general an order of magnitude greater than the number of monopoles, i.e.,
$\rho_{disc} \simeq 10 \rho_{monop}$. In spite of this order of magnitude
difference in densities, we see from Eq. (\ref{strucdefect}) that the monopoles
and disclinations should have comparable contributions to the structure factor
in the range of $k$ values used in our plots. Thus, it is not clear why we
obtain an exponent between 4 and 5, though one possibility is that we are
seeing two-dimensional effects due to the anisotropic shape of our MD cell
(recall that point disclinations in a two-dimensional nematic should yield a
power law of 4). It is also possible that we have overestimated the number of
monopoles, whose contribution to the structure factor would yield an exponent
greater than 5. As discussed above in Sec.  \ref{sec:coarsening}, the monopoles
fluctuated quite rapidly, and it is possible that some apparent monopoles that
we identified using the algorithm described in Sec.  \ref{sec:algorithm} are
not in fact topological defects.

\par
The crossover value of $k$ separating the defect dominated and thermal
fluctuation dominated regimes is of the order of magnitude predicted by Eq.
(\ref {strucdefect}), namely, $k \sim (2 \pi^3 K \rho_{disc}/ k_B T)^{1/3}\sim
(24 \pi^4 \rho_{monop}/k_B T)^{1/4}$. This crossover value decreases with time,
as can be seen by comparing Fig. \ref{strucfig27} with the later time data of
Fig. \ref{strucfig40}. The crossover value of $k$ in the latter figure is about
half of the corresponding value in the former figure, consistent with the
relative densities of defects at the two times.
\par
\begin{figure}\center

\epsfxsize=12cm \epsffile{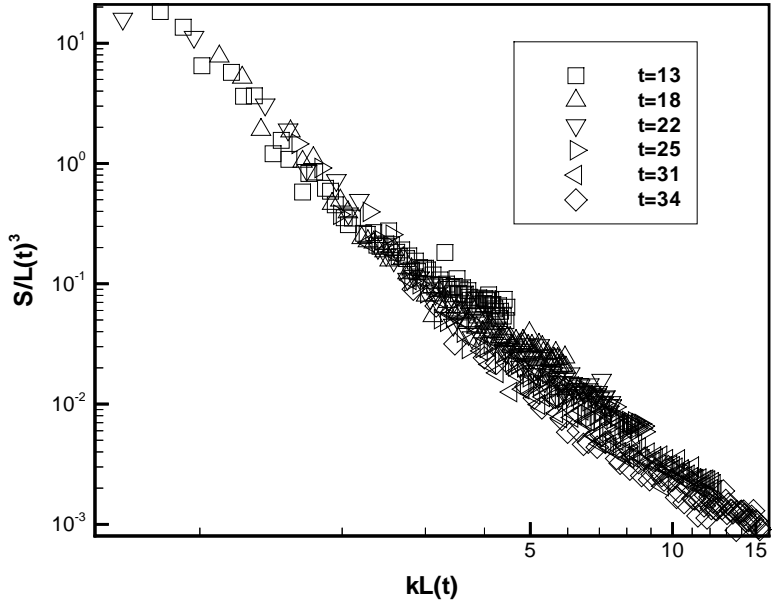}
\caption{Plot of the dynamical scaling function $g=S(k,t)/(L(t))^3$ as a
function of the scaling variable $kL(t)$, (see Eq. (\ref{strucscaling})). Note
the clear breakdown of scaling for large values of $kL(t)$.}
\label{scaledstrucfac}
\end{figure}

As discussed in ref. \cite{zap-gold1} we would expect the crossover to the
thermal fluctuation dominated regime to be accompanied by a breakdown of the
dynamical scaling hypothesis because it  assumes that thermal fluctuations play
no role in the behavior of real--space or Fourier--space correlation functions.
To test this expectation we used our data for $S(k,t)$ during the time range
spanning the coarsening process to plot the scaling function $g$ defined in Eq.
 (\ref{strucscaling}). This plot is shown in Fig. \ref{scaledstrucfac}, where
we clearly see the breakdown of scaling  for $kL(t)$ greater than approximately
4 or 5, corresponding to values of $r/L(t)$ less than approximately one. Note
that the numerical range of $g$ is much larger than the range of $f$, the
corresponding scaling function for $C(r,t)$ (Eq. (\ref{corrscaling}) and Fig.
\ref{corrsfig}), so that the breakdown in scaling is easier to see in the
structure factor data. The wider horizontal range for $kL(t)$ compared to
$r/L(t)$ also makes the breakdown clearer.

\par
 Note in our plots, $k \geq R^{-1}$, for nearly all values of the disclination
loop radius $R$. Thus, in this regime we expect to see a power--law
contribution to $S(k)$ from the twist dislination loops. With a simulation of a
larger system it might be possible to study the regime $k \ll R^{-1}$ where the
twist disclination loops are expected to make no contribution to $S(k)$, as
discussed in Sec. \ref{sec:theory}.
\section{Conclusions}
In conclusion, we have shown that the \gb\ potential is fruitful
for studying the behavior of the wide variety of \tds\ generated in a quench
from the isotropic to the nematic phase. At least for the \gb\ parameters
chosen here,
twist \disc\ loops were the dominant defects, and we did not, aside from
possibly one isolated line, observe dynamically generated wedge disclination
loops. This result, if it is not an artifact of our relatively small system
size, has important implications for the interpretation of scattering
experiments on quenched nematics, following upon the ideas of Ref.
\cite{zap-gold1}. As we discussed in Sec. II measurements of the structure
factor show scaling with an exponent between 5 and 6. Twist disclination lines
yield an exponent of 5, whereas monopoles (which are relatively few in number
in experimental systems) and wedge disclination lines (which are not
energetically favorable) yield 6. Thus, it remains an open question as to why
the exponent observed is greater than 5 over some appreciable range of
wavevector.
\par
Our computed real--space correlation function exhibited good dynamical scaling
over the limited range of distances available, though the structure factor
appears to be provide a more sensitive test of scaling. In our structure factor
data we could clearly see the breakdown of dynamical scaling and the crossover
to the thermal fluctuation dominated behavior, in accord with the predictions
of Zapotocky and Goldbart\cite{zap-gold1}. Clearly, simulations of even larger
\gb\ systems would be of interest to further address the issues raised here.
\section*{Acknowledgements}
Helpful discussions with Prof. G. Crawford are gratefully acknowledged.
Computational work in support of this research was performed at the Theoretical
Physics Computing Facility at Brown University and at the San Diego
Supercomputing Center under the auspices of the National Partnership for
Advanced Computational Infrastructure.  This work was supported by the National
Science Foundation under grants nos. DMR-9528092 and DMR98-73849.

\end{document}